\documentclass[12pt,letterpaper]{article}

\usepackage{amsmath,amssymb,calc}
\usepackage{graphicx}

\newcommand\be{\begin{equation}}
\newcommand\ee{\end{equation}}
\newcommand{\bea}{\begin{eqnarray}}
\newcommand{\eea}{\end{eqnarray}}
\newcommand{\half}{\frac{1}{2}}

\newcommand{\p}{\partial}
\newcommand{\diff}{{d}}
\newcommand{\ra}{\rightarrow}
\newcommand{\Mcal}{{\mathcal M}}
\newcommand{\Ncal}{{\mathcal N}}
\newcommand{\tOmega}{\tilde{\Omega}}
\newcommand{\btOmega}{\overline{\tilde{\Omega}}}

\newcommand{\nn}{\nonumber}

\def\id{\protect{{1 \kern-.28em {\rm l}}}}

\def\cW{{\cal W}}
\def\ve{\varepsilon}
\def\tJ{{\tilde{J}}}
\def\tO{{\tilde{\Omega}}}

\font\mybb=msbm10 at 12pt

\def\bb#1{\hbox{\mybb#1}}

\def\id{\protect{{1 \kern-.28em {\rm l}}}}

\begin{document}

\begin{titlepage}
\begin{center}
\hfill QMUL-PH-06-01 \\
\hfill {\tt hep-th/0602039}\\
\vskip 15mm

{\Large {\bf Flux Superpotential in Heterotic M--theory \\[3mm] }}

\vskip 10mm

{\bf Lilia Anguelova and Konstantinos Zoubos}

\vskip 4mm
{\em Department of Physics, Queen Mary, University of London}\\ 
{\em Mile End Road, London E1 4NS, United Kingdom} \\
{\tt l.anguelova@qmul.ac.uk, k.zoubos@qmul.ac.uk}\\

\vskip 6mm

\end{center}

\vskip .1in

\begin{center} {\bf ABSTRACT }\end{center}

\begin{quotation}\noindent

We derive the most general flux--induced superpotential for ${\cal N}=1$ M--theory compactifications on seven--dimensional manifolds with $SU(3)$ structure. Imposing the appropriate boundary conditions, this result applies for heterotic M--theory. It is crucial for the latter to consider $SU(3)$ and not $G_2$ group structure on the seven--dimensional internal space. For a particular background that differs from $CY(3) \times S^1\!/\mathbb{Z}_2$ only by warp factors, we investigate the flux--generated scalar potential as a function of the orbifold length. We find a positive cosmological constant  minimum, however at an undesirably large value of this length. Hence the flux superpotential alone is not enough to stabilize the orbifold length at a de Sitter vacuum. But it does modify substantially the interplay between the previously studied non--perturbative effects, possibly reducing the significance of open membrane instantons while underlining the importance of gaugino condensation. 

\end{quotation}
\vfill

\end{titlepage}

\eject

\tableofcontents

\section{Introduction}

The four--dimensional effective action of string and M--theory compactifications contains many moduli --- scalar fields arising from deformations of the internal space. The lack of a potential for these fields leads to vacuum degeneracy and loss of predictability of the four-dimensional coupling constants. The resolution of this longstanding moduli--stabilization problem has started taking shape only in recent years. In particular, it was shown that the superpotential, generated by turning on background fluxes, stabilizes the complex structure moduli in type IIB \cite{GKP}\footnote{The K\"{a}hler moduli can be stabilized by non--perturbative effects \cite{KKLT}.} and all geometric moduli in type IIA \cite{DKPZ,VZ,DGKT}. It was also argued that flux compactifications can fix the radial modulus in the heterotic string case \cite{BBDP}. 

The form of the flux superpotentials in string and M--theory was first deduced in \cite{GVW, G, BG} and subsequently refined and verified in \cite{BW, BC} and many others. For a recent thorough review of the literature on flux compactifications and an exhaustive list of references see \cite{MG}. Remarkably, heterotic M--theory has remained aside from these considerations despite the fact that it admits only nonzero background--flux solutions. The reason is that its action is mostly unknown beyond order $\kappa^{2/3}$, where $\kappa$ is the gravitational coupling constant. Recall that this theory is obtained by compactifying M--theory on a $CY(3)$ times an interval $S^1/\bb{Z}_2$. The presence of boundary sources leads to nonvanishing 11d supergravity four--form $G$ and an effective action which has an expansion in powers of $\kappa^{2/3}$ \cite{HW,HW2}. Due to technical difficulties though, only the bosonic terms of ${\cal O}(\kappa^{2/3})$ together with some ${\cal O}(\kappa^{4/3})$ terms have been found so far \cite{LOW}. On the other hand, the flux superpotential is at least linear in the flux and since the background G-flux is itself of order $\kappa^{2/3}$, then the flux--superpotential contribution to the bosonic part of the effective action would be of order $\kappa^{4/3}$ and higher (as the potential is schematically $U\sim W^2$). So the currently known results from dimensional reduction of the bosonic 11d supergravity action \cite{LOW} are not enough to detect the presence of a flux-generated superpotential. 

However, one can extract the superpotential $W$, without resolving the difficulties of going to higher orders in the heterotic M--theory action\footnote{These difficulties may eventually be overcome in the approach of \cite{IM,IM2}.}, by using the fact that $W$ appears linearly in the gravitino mass term of the 4d effective theory. This strategy for obtaining the superpotential was already used in the cases of the heterotic string on half-flat manifolds \cite{GLM}, M--theory on $G_2$ structure manifolds \cite{HM} and (massive) IIA on 
$SU(3)$ structure ones \cite{HP}. We will apply it to ${\cal N}=1$ M--theory compactifications on $SU(3)$ structure spaces. Imposing the appropriate boundary conditions, these considerations become relevant for heterotic M--theory. 

The importance of group structures for the study of flux compactifications was realized since the work of \cite{GMPW}. More precisely, this is the suitable generalization of the notion of holonomy of the internal space for manifolds with torsion, the latter being induced by the backreaction of the flux on the geometry. For the case of $SU(3)$ structure, there are two Majorana spinors which are covariantly constant with respect to the connection with torsion, determined by the flux. Nevertheless, the four-dimensional effective theory can have either $\Ncal=1$ or $\Ncal=2$ supersymmetry depending on the ansatz giving the eleven--dimensional spinors in terms of external and internal ones. Our derivation of the superpotential is performed for the most general $\Ncal=1$ ansatz, given in \cite{BCL}.\footnote{We comment on the compatibility of this ansatz with the heterotic M-theory boundary conditions in Subsection \ref{Prelim}.} The latter work extracts partial information about the superpotential with on-shell methods and our results are compatible with theirs.

Obtaining the superpotential from dimensional reduction of the fermionic terms of the eleven--dimensional supergravity action, gives a result which is exact in terms of the $\kappa^{2/3}$ expansion of heterotic M--theory. One can expand it to any desired order, but it is guaranteed (as we already explained) that the lowest order in the scalar potential that it  contributes to is $\kappa^{4/3}$. Hence one may wonder whether it is justified to keep this flux--induced superpotential, $W_{flux}$, in the scalar potential given that most of the bosonic part of the heterotic M--theory effective action is not known at the orders at which $W_{flux}$ matters. However, the recent exciting advances in understanding the strongly coupled limit of the $E_8 \times E_8$ heterotic string, namely finding de Sitter \cite{BCK} and assisted inflation \cite{BBK} solutions, employed in an essential way non--perturbative effects (to generate $W_{non-pert}$) and an exact, containing all orders in $\kappa^{2/3}$, background \cite{CK}. Therefore, it is a perfectly legitimate and even pressing question to try to take into account any corrections to their considerations, that one can estimate.

We find that for a background that differs from the zeroth order one, $CY(3) \times S^1/\bb{Z}_2$, only by warp factors, as in \cite{CK}, the flux-generated superpotential modifies substantially the analysis of the orbifold length stabilization. More precisely, the scalar potential due to $W_{flux}$ alone, without the inclusion of any non--perturbative effects, does have a de Sitter minimum. However, it occurs at a value of the orbifold length that is slightly larger than the maximal allowed one. Therefore, it is of crucial importance to also take into account gaugino condensation, which seems poised to help in lowering the location of the minimum. 
 On the other hand, the role of membrane instantons does not seem significant once the flux superpotential is included. For more general backgrounds, in which the initial $CY(3)$ is deformed to a non--K\"{a}hler manifold, the form of $W_{flux}$ is such that potentially it could stabilize all geometric moduli, similarly to type IIA. 

We also find that, when $W_{flux}$ is non-vanishing, the charged matter fields $C^I$, originating from the boundary $E_8$ gauge multiplets, are not stabilized at exponentially suppressed values anymore. Hence the question whether or not one can neglect them while extremizing the potential w.r.t. to the rest of the fields, as in \cite{BCK}, is more subtle and its answer depends on the explicit backgrounds.

The organization of this paper is the following. In Section 2 we derive  the flux--induced superpotential from dimensional reduction of the fermionic 11d terms and reading off the coefficient of the resulting 4d gravitino mass term. In Subsection 2.1 we review necessary material about $SU(3)$ structure manifolds and in Subsection 2.2 we perform the computation of the superpotential. We also make the observation that for the application to heterotic M--theory it is of crucial importance to consider $SU(3)$ structure on the internal seven--dimensional manifold rather than $G_2$ structure. In Section 3 we consider the implications of this superpotential for moduli stabilization. More precisely, for a particular heterotic M--theory background we investigate whether the orbifold length can be stabilized at a dS vacuum without the need of non--perturbative effects. As an aside we also derive the modification of the generalized Hitchin flow equations due to the eleven--dimensional $R^4$ term to order $\kappa^{4/3}$. In Section 4 we discuss the role of perturbative corrections to the K\"{a}hler potential of the universal moduli and also the influence of $W_{flux}$ on the vevs of the charged matter fields $C^I$. Finally, in the Appendix 
we summarize our conventions and give some results on seven--dimensional $SU(3)$ structures that are necessary for our calculations.

\section{Flux superpotential}

Before embarking on the computation of the flux--induced superpotential, let us explain the relevance of $SU(3)$ structures for our considerations.

At zeroth order the background of Ho\v{r}ava--Witten theory, which leads to $\Ncal=1$ supersymmetry in four dimensions, 
has the form $\bb{R}^{1,3}\times CY(3) \times S^1/\bb{Z}_2$. However, it is well-known that the presence of boundary sources modifies the Bianchi identity of the 11d supergravity 4-form field strength $G$. As a result, any solution of this theory must have non-vanishing flux. The deformation, due to the background flux, 
appears at orders $\kappa^{2/3}$ and higher. For special choices of the flux, the backreaction on the geometry 
can be encoded only in warp factors. But for generic $G$-flux, the appearance of warp factors has to be accompanied 
by a deformation of the initial $CY(3)$ to a non--K\"{a}hler manifold \cite{EW,CK}. As the dependence of the latter 
six-dimensional space on the eleventh (interval) direction is one of the characteristic features of the strongly coupled 
limit of the $E_8 \times E_8$ heterotic string, one obtains the picture that the internal seven--dimensional manifold 
of a Ho\v{r}ava--Witten compactification is a fibration of a non--K\"{a}hler manifold along an interval. Such an internal 
space can be described in the language of group structures as a seven-manifold with $SU(3)$ structure. The weak 
coupling limit, which is the $E_8\times E_8$ string on a six-dimensional space with $SU(3)$ structure, is achieved by 
turning off the dependence on the interval.

One might have thought that since the 4d Poincar\'e--invariant effective theory has $\Ncal=1$ supersymmetry, the internal space
 could be described most usefully in terms of $G_2$ structure. However, it was shown in \cite{BJ} that $G_2$ structure does 
not allow nontrivial fluxes and warp-factors.\footnote{The only nontrivial, i.e. different from $G_2$ holonomy, situation 
allowed by $G_2$ structure considerations is compactification on a {\it weak} $G_2$ manifold with the external space being 
$AdS_4$ \cite{BJ}.} On the 
other hand, $SU(3)$ structure can lead to either $\Ncal=1$ or $\Ncal=2$ supersymmetry depending on the ansatz for the 
11d supergravity Killing spinor. We will be more precise on the ansatz of interest to us (i.e. for the ${\cal N}=1$ case) in a short while.

Let us now recall the basic features of $SU(3)$ structure manifolds that we will need. 
For more details we refer the reader to e.g. \cite{CS}.

\subsection{$SU(3)$ structures in 6 and 7 dimensions}

In six dimensions an $SU(3)$ structure is 
defined by a real 2-form $J$ and a holomorphic 3-form $\Omega$ that satisfy the 
compatibility conditions:
\be \label{compat}
J \wedge \Omega = 0 \qquad {\rm and} \qquad J \wedge J \wedge J = \frac{3i}{4} \Omega \wedge \bar{\Omega} \, .
\ee
Their non-closedness characterizes the deviation from $SU(3)$ holonomy:
\bea \label{tor}
dJ &=& -\frac{3}{2} Im (\cW_1 \bar{\Omega}) + \cW_3 + J \wedge \cW_4 \nn \\
d \Omega &=& \cW_1 J\wedge J + J \wedge \cW_2 + \Omega \wedge \cW_5
\, , \eea where $\cW_i$, $i=1,...,5$, are called torsion classes. They have to satisfy 
the following relations: 
\be \label{rel}
J\wedge \cW_3=0 \, ,\qquad J\wedge J\wedge \cW_2=0 \, ,\qquad \Omega\wedge \cW_3=0 \, .
\end{equation}
The class $\cW_1$ appears as the coefficient of both $dJ^{(3,0)\oplus (0,3)}$ and $d\Omega^{(2,2)}$ due to (\ref{compat}) and (\ref{rel}). 
In weakly coupled heterotic string compactifications, supersymmetry
requires that the internal manifolds have $\cW_1 = 0 = \cW_2$
\cite{CCDLMZ}, which, in particular, means that they are complex
although generically non--K\"{a}hler. 

A seven--dimensional $SU(3)$ structure is defined by a real vector $v$,\footnote{For convenience we
will denote its dual one-form by the same letter.} a real 2-form
$\tilde{J}$ and a holomorphic 3-form $\tilde{\Omega}$, with $\tilde{J}$ and
$\tilde{\Omega}$ satisfying (\ref{compat}) and in addition: 
\be \label{compv} 
i_v \tilde{J} = 0 \qquad {\rm and} \qquad i_v
\tilde{\Omega} = 0 \, . 
\ee 
The exterior differentials $dv$, $d\tilde{J}$ and $d\tilde{\Omega}$ now determine 14 torsion classes. For our purposes though, it is not necessary to write down their explicit form; more details on them can be found in e.g. \cite{DP,BCL}. It is important to note 
that due to (\ref{compv}) any seven--dimensional $SU(3)$ structure manifold naturally is a fibration of a 6d 
$SU(3)$ structure space over an additional dimension. Furthermore, any compact 7d orientable manifold admits 
two nowhere vanishing vector fields \cite{ET}, which implies that every compact 7d spin manifold admits $SU(2)$ structure \cite{FKMS}. As a result,
it also admits $SU(3)$ and $G_2$ structure. So we have not imposed any restriction (other than being spin, which 
is anyway necessary for supergravity) on the internal manifold by requiring that it has $SU(3)$ structure. However, 
physics does depend on which G--structure the reduction of the eleven--dimensional spinors is adapted to, as will become 
more clear below.

\subsection{The superpotential} \label{Superpotential}

Now, we will extract the superpotential from the gravitino mass term of the 4d effective theory. 
This term follows straightforwardly from dimensional reduction of the fermionic terms of the 11d action. As already mentioned, the same 
approach to obtaining the superpotential in a flux compactification has also been used in \cite{GLM} for
the weakly coupled heterotic string on half--flat (a special case of $SU(3)$ structure) manifolds, in 
\cite{HP} for IIA with $SU(3)$ structure, and (more relevant to our case) 
in \cite{HM} for M--theory on $G_2$--structure manifolds. 

Let us also note, that partial results on the superpotential relevant to our case have been obtained in \cite{BCL}.
However, their consideration is intrinsically on-shell and hence it cannot distinguish between 
fluxes and torsions. Besides, it can only give information about the value of the superpotential at a
given vacuum. On the other hand, we are interested in the superpotential as a part of the effective action. This is an off--shell quantity that contains information about the independent
fluctuations of the flux and geometry degrees of freedom. For this reason we want to obtain it via dimensional 
reduction of the eleven--dimensional supergravity action. We will see that our results are compatible
with the partial information found in \cite{BCL}.

\subsubsection{Preliminaries} \label{Prelim}

The Lagrangian for supergravity in 11 dimensions \cite{CJS,CJ} can be written as:\footnote{We have made the following rescalings relative to \cite{CJS}:
$\kappa\ra \kappa/\sqrt{2},G_{(4)}\ra  G_{(4)}/(\sqrt{2}\kappa),\Psi\ra\Psi/\kappa$. We 
use the mostly--plus convention for the metric
and, since (as we will discuss) we use a real representation for the gamma matrices, we take 
$\Gamma\ra i\Gamma$. This form of the action matches that of \cite{HM}, who considered
the $G_2$ structure case, and thus allows an easier comparison with their results.}
\be
\begin{split} \label{Supergravity}
\mathcal{L}= \frac{1}{\kappa^2} &\left[\frac{e}{2}R-\frac{e}{2}\overline{\Psi}_M\Gamma^{MNP}D_N\Psi_P
-\frac{1}{4}G\wedge*G-\frac{1}{12}C\wedge G\wedge G \right. \\
&\left. 
-\frac{e}{192}(\overline{\Psi}_M\Gamma^{MNPQRS}\Psi_N+12\overline{\Psi}{}^P\Gamma^{QR}\Psi^S)
G_{PQRS}
\right]
\end{split}
\ee
where $G=dC$ and $\overline{\Psi}=\Psi^\dagger\Gamma_0$. 
The full action includes four--fermion terms (including those hidden in the precise 
definition of the spin connection)
but they are not relevant for obtaining the gravitino mass term in four dimensions 
(which is all we will require) and so we do not exhibit them. In the  heterotic M--theory action there are also boundary terms containing the $E_8$ gauge fields; we will comment on their implications later on.

To perform the dimensional reduction we need an ansatz for the
embedding of the 4d gravitino, $\psi_\mu$, in the 11d one, $\Psi_{M}$, with the
help of the internal spinors. The most general ansatz, which leads
to an $\Ncal=1$ compactification on an $SU(3)$ structure space, is the one
considered in \cite{BCL}:\footnote{It was remarked in \cite{BCL} that the ansatz (\ref{grav}) cannot be embedded in Ho\v{r}ava-Witten theory because it does not survive the $\mathbb{Z}_2$ projection, since it does not give a 6d chiral spinor in the internal space orthogonal to the interval. However, the internal space of heterotic M-theory is $CY(3)\times S^1/\mathbb{Z}_2$ only at leading order. At higher orders, as already explained, it is in principle a seven-dimensional manifold with  $SU(3)$ structure. And one only has to ensure that the spinors are chiral at the boundaries (but not necessarily in the bulk), which is achieved by imposing the boundary condition $a_L \rightarrow 0$ or $a_R \rightarrow 0$ as one approaches the boundary. For a related discussion, see for example \cite{LS}.} 
\be \label{grav} 
\Psi_{\mu} = \psi_{\mu} \otimes (a_L \theta + a_R^* \theta^*) + \psi_{\mu}^* \otimes (a_L^* \theta^* + a_R \theta) \, , 
\ee 
where $a_L$ and $a_R$ are complex
functions of the internal coordinates, $\psi_{\mu}$ is a 4d Weyl
gravitino and $\theta$, $\theta^*$ are the $SU(3)$ singlet internal
spinors.\footnote{To be more precise, $\theta = \epsilon_1+ i
\epsilon_2$ with the Majorana spinors $\epsilon_{1,2}$ being the
$SU(3)$ singlets.} We will normalize $\theta^\dagger\theta=\theta^T\theta^*=1$. Note
that we also have $\theta^T\theta=\theta^\dagger\theta^*=0$. 
Taking $a_L = 0$ or $a_R = 0$, one obtains the
ansatz that was studied in \cite{DP,BJ}. On the other hand, taking
$a_L = a_R^*$ one recovers the ansatz appropriate for
compactification on $G_2$ structure \cite{HM,BJ2}. 

We also need to know the relations between the $SU(3)$ structure defining forms $v$, $\tilde{J}$, $\tilde{\Omega}$ and the non-vanishing spinor bilinears that one can construct from $\theta$ and $\theta^*$. As nicely
summarized in \cite{BCL}, these are given by (we write down only the
relations that will be necessary for our computation): 
\be\label{spbilin}
\begin{split}
 v_a = \theta^{\dagger} \gamma_a \,\theta \, , \qquad
\tJ_{ab} =& -i \,\theta^{\dagger} \gamma_{ab} \,\theta \, , \qquad
\tO_{abc} = -i \,\theta^T \gamma_{abc} \,\theta \, ,\\ 
(v\wedge\tJ)_{abc} = -i \,\theta^{\dagger} \gamma_{abc} \,\theta \, , \quad
(\tJ\wedge \tJ)_{abcd} &= \,-2\theta^{\dagger} \gamma_{abcd} \,\theta\, , \quad 
(v\wedge \tO)_{abcd} = i \,\theta^T \gamma_{abcd}\,\theta \, , 
\end{split}
\ee
 where $a,b,... = 1,...,7$ are the internal
indices. We use the conventions of \cite{BCL}, in which $\gamma_a$
are purely imaginary and hermitian. The (real) eleven--dimensional gamma
matrices are given in terms of the four-dimensional, $\gamma^{\mu}$,
and seven--dimensional, $\gamma^a$, gamma matrices by $\Gamma_{\mu} = \gamma_{\mu}
\otimes 1\!\!1$ and $\Gamma_{a+4} = \gamma \otimes \gamma_a$ with
$\gamma = \frac{i}{4!} \ve_{\mu \nu \rho \sigma} \gamma^{\mu}
\gamma^{\nu} \gamma^{\rho} \gamma^{\sigma}$. We also define $\Gamma_0\equiv\Gamma_4$. 

Since in four dimensions the gravitino mass term has the form
\footnote{Note that this is the correct form of the spinor bilinear in the mass term of a 
4d Weyl (as opposed to Majorana) gravitino.} 
\be \label{Mass}
\mathcal{L}_{\frac{3}{2}}^{(mass)} 
=\frac{1}{2} e^{K/2} (W \bar{\psi}_{\mu} \gamma^{\mu \nu} \psi_{\nu}^* + c.c.) \,, 
\ee 
we only need to extract the coefficient of the terms
containing $\bar{\psi}_{\mu} \gamma^{\mu \nu} \psi_{\nu}^*$.
 For simplicity and easier comparison with
existing literature, we start by considering a direct product metric
\be \label{directprod}
ds^2_{11} = ds^2_{4}+ds^2_{7} 
\ee
and will include the appropriate
warp factors later on. Let us also note that the component $\Psi_a = \psi_a \otimes (a_L \theta + a_R^* \theta^*) + \psi_a^* \otimes (a_L^* \theta^* + a_R \theta)$ will lead to mixed terms between the gravitino $\psi_{\mu}$ and the spin 1/2 fermion $\psi_a$. Hence, to obtain the standard 4d supergravity, one has to redefine the gravitino by a shift proportional to $\psi_a$. However, this does not affect the mass term we are interested in and so for convenience we will work with the original $\psi_{\mu}$.

It is clear that from (\ref{Mass}) one can only extract the combination $e^{K/2} W$. To address its splitting 
into a K\"{a}hler potential and a superpotential, we need to be more
careful with the normalization of the various terms in the effective
action that results from the dimensional reduction. In particular, in order to obtain the canonical normalization of
the kinetic terms in the 4d supergravity action one has to perform the following rescalings
\cite{CJ}: 
\be \label{res}
g_{\mu\nu} \rightarrow {\cal V}^{-1}_{(7)} g_{\mu \nu} \, , \qquad
\gamma_{\mu} \rightarrow {\cal V}^{-1/2}_{(7)} \gamma_{\mu} \, ,
\qquad \psi_{\mu} \rightarrow {\cal V}^{-1/4}_{(7)} \psi_{\mu} \, ,
\ee 
where ${\cal V}_{(7)}$ is the volume of the seven--dimensional
internal space. In addition, given that $\theta^\dagger\theta=1$, 
the normalization of the gravitino kinetic term forces
$|a_L|^2+|a_R|^2=1$. Examining the effect of the rescalings of the 4d fields in (\ref{res})
on the gravitino mass term,  we are led to the result:
\be 
e^{K/2} W = \frac{1}{{\cal V}_{(7)}^{3/2}} \,[{\rm result
\,\, of \,\, dim. \, reduction}] \, . 
\ee 
Now, in M--theory compactifications on $G_2$ holonomy spaces the K\"{a}hler 
potential turns out to be 
\be \label{Ka}
K = - 3 \ln {\cal V}_{(7)} \, . 
\ee
This formula was found to give consistent results for compactifications on $G_2$ structure manifolds as well \cite{HM}. It is natural then to assume that it also applies
for compactifications on seven--dimensional
spaces with $SU(3)$ structure. Hence the superpotential $W$ can be read off immediately from the 4d gravitino mass term.\footnote{One might have thought that 
in heterotic M--theory things are more subtle, since the K\"{a}hler potential for the universal moduli, at leading order in the $\kappa^{2/3}$ expansion, is known
to be \cite{LOW}: 
$K_0 = -\ln (S + \bar{S}) - 3\ln (T + \bar{T})$, 
where $\mathrm{Re}S = {\cal V}$ and $\mathrm{Re}T = {\cal L} {\cal V}^{1/3}$ with ${\cal V}$
being the volume of the six-dimensional non--K\"{a}hler manifold and
${\cal L}$ the length of the interval. Hence the power of ${\cal V}_{(7)}$, resulting from the rescaling (\ref{res}), cannot be completely absorbed in the K\"{a}hler potential $K_0$ since $e^{K_0/2}$ equals ${\cal V}^{1/2} {\cal V}^{-3/2}_{(7)}$ instead of ${\cal V}^{-3/2}_{(7)}$. Note however, that there are many other moduli in addition to the universal ones and so $K_0$ is not the full answer for the K\"{a}hler potential.} For convenience, from now on we will work with the rescaled fields and view the power of ${\cal V}_{(7)}$ as already being taken care of via (\ref{Ka}).

We are finally ready to do the dimensional reduction.

\subsubsection{Reduction of the flux term}

Using (\ref{grav}), we find from the 11d flux
term: 
\bea \label{flt}
\frac{-1}{8 \times4!} \bar{\Psi}_M \Gamma^{MNPQRS} \Psi_N G_{PQRS} &=& 
\frac{-1}{8\times 4!}\bar{\psi}_{\mu} \gamma^{\mu \nu} \psi_{\nu}^* 
\left[ a_L^* a_R \theta^{\dagger} \gamma^{abcd} \theta 
+ (a_L^*)^2 \theta^{\dagger} \gamma^{abcd} \theta^* \right. \nn \\
&+& \left. a_R a_L^* \theta^T \gamma^{abcd} \theta^* + (a_R)^2
\theta^T \gamma^{abcd} \theta \right] G_{abcd} + ... \, .
\eea 
The ``...'' denote terms that do not contain $\bar{\psi}_{\mu} \gamma^{\mu \nu}
\psi_{\nu}^*$. Relations (\ref{spbilin}) imply, upon using that
$*_7(\tJ \wedge \tJ) = 2 \tJ \wedge v$ and $*_7 (\tO \wedge v) = i
\tO$, that the part of the superpotential extracted from
(\ref{flt}) can be written as 
\be \label{Wfl} 
W^{(flux)} = \frac{1}{2} \int a_L^* a_R \,\,G \wedge \tJ \wedge v + \frac{1}{4} \int \left[ (a_R)^2 \,\,G\wedge \tO + (a_L^*)^2 \,\,G \wedge \btOmega \right]
\, . 
\ee 
This is compatible with \cite{BCL} (modulo the redefinitions: $a_L^* \rightarrow a_L$ and $a_R \rightarrow a_R^*$). The constraint that the superpotential vanish for
$a_L=0$ in their on-shell approach just means that $W^{(flux)} (a_L
=0)$ has only Minkowski minima but not $AdS_4$ ones. The expression
(\ref{Wfl}) describes a contribution to the flux superpotential for {\it generic} ${\cal N}=1$ compactifications of M--theory on a 7d manifold with $SU(3)$
structure. In order for it to be applicable to heterotic M--theory,
the equations of motion must be compatible with the appropriate boundary conditions, and, in particular, the condition that the function $a_L$ tends to zero as one approaches the visible boundary. If such a solution is not
possible, then one should set $a_L = 0$ in the ansatz (\ref{grav}),
to be able to recover the well--known superpotential $\int H\wedge
\tO$ \cite{BG,BC} in the weakly coupled limit, where the NS--NS flux 
is $H = i_v G$. Notice that the first term in (\ref{Wfl}) drops out in 
the weakly coupled limit even when both $a_L, a_R \neq 0$ as in that limit 
$G =v \wedge H$.

\subsubsection{Reduction of the kinetic term}

Let us turn now to computing the contribution from the 11d gravitino kinetic term. 
This will lead to new terms in the superpotential compared to the results of \cite{BCL}. 
Some relations that will be useful for the calculation are summarized in appendix \ref{Relations}.
Consider:
\bea \label{kin}
-\frac{1}{2} \bar{\Psi}_M \Gamma^{MNP} D_N \Psi_P &=& 
\frac{1}{2} \bar{\psi}_{\mu} \gamma^{\mu \nu} \gamma \psi_{\nu}^* 
\left[ (a_L^*)^2 \theta^{\dagger} \gamma^a \nabla_a \theta^* 
+ a_L^* a_R (\theta^{\dagger} \gamma^a \nabla_a \theta 
+ \theta^T \gamma^a \nabla_a \theta^*) \right. \nn \\
&+& \left. (a_R)^2 \theta^T \gamma^a \nabla_a \theta 
+ v^a (a_L^* \partial_a a_R - a_R \partial_a a_L^*) \right] + ... \, .
\eea
To obtain a mass term of the form (\ref{Mass}), we utilize the chirality properties of
$\psi_\mu$, i.e. $\gamma\psi_\mu^*=-\psi_\mu^*$. To get rid of the derivatives we use that
\be \label{nab}
\nabla_a \theta = \frac{1}{4} \tau_{abc} \gamma^{bc} \theta \, ,
\ee
where $\tau_{abc}$ is the geometric (con)torsion of the internal manifold. Hence, making use of (\ref{spbilin}), we obtain for the geometric part of the superpotential density the following:
\be \label{wg}
w^{(geom)} = -\frac{i}{4} \left[2 a_L^* a_R  (v\wedge \tJ)^{abc} \tau_{abc} 
+  (a_L^*)^2 \btOmega{}^{abc} \tau_{abc} 
+ a_R^2 \tO^{abc} \tau_{abc}\right] 
- v^a (a_L^* \partial_a a_R - a_R \partial_a a_L^*).
\ee
To write the superpotential $W^{(geom)} = \int \sqrt{g} \,w^{(geom)}$ in form 
language we need to use the relations (see appendix \ref{Relations} for more details)
\be \label{dJ}
(dv)_{ab} = 2 \tau_{[ab]}{}^c v_c \, , \qquad (d\tJ)_{abc} = 6 \tau_{[ab}{}^d \tJ_{|d|c]} \, , \qquad (d\tO)_{abcd} = 12 \tau_{[ab}{}^e \tO_{|e|cd]} \, ,
\ee
which follow from $v$, $\tJ$ and $\tO$ being $SU(3)$ invariant forms .
Together with $\tJ_a{}^b \tJ_b{}^c = -\delta_a^c + v_a v^c$, equations (\ref{dJ}) 
allow one to express the various components of the torsion $\tau_{abc}$ in terms of $d\tJ$, $\tJ$, $d\tO$, $\tO$, $dv$ and $v$. 
As a consequence we obtain, upon using $\tO^{abc} \tJ_c{}^d = -i \tO^{abd}$ and $*_7 \tO = -i \tO \wedge v$ together with
(\ref{compv}),\footnote{As usual, we have normalized $v^a v_a =1$.} that
\bea \label{Wgeom}
W^{(geom)} = \int \sqrt{g} \,w^{(geom)} &=& 
\frac{i}{4} \int (a_R)^2 \,v \wedge d\tJ \wedge \tO 
+ \frac{i}{4} \int (a_L^*)^2 \,v \wedge d\tJ \wedge \btOmega \nn \\
&-& \frac{i}{16} \int a_L^* a_R [d\tO \wedge \btOmega 
+ d\btOmega \wedge \tO + 4 dv\wedge v \wedge \tJ \wedge \tJ] \nn \\
&-& \frac{1}{6} \int (a_L^* d a_R - a_R d a_L^*) \wedge \tJ \wedge \tJ \wedge
\tJ \, . 
\eea 
One might have thought that the first term in (\ref{wg}) would also lead to a contribution 
to $W^{(geom)}$ of the form $d\tJ \wedge \tJ \wedge \tJ$, but this does not happen because 
the second expression in (\ref{dJ}) applied to $v^a\tJ^{bc} \tau_{abc}$ yields 
(due to (\ref{compv})) $v^a (d\tJ)_{ab}{}^b+v^a (d\tJ)_{abc} v^bv^c= 0$. 
Note that the last term in (\ref{Wgeom}) could as well have been written, up to a 
numerical coefficient, as $\int (a_R da_L^* - a_L^* d a_R) \wedge \tO \wedge \btOmega$, 
but due to (\ref{compat}) these two expressions are equivalent. Again, if we
set $a_L = 0$ we can easily recover the superpotential term $d\tJ
\wedge \tO$ that is known to be present in compactifications of the weakly coupled 
heterotic string on non--K\"{a}hler manifolds \cite{BBDP,CCDL}. As before,
$W^{(geom)}$ is valid for any ${\cal N}=1$ compactification of M--theory on a
seven--dimensional manifold with $SU(3)$ structure. If one wants to
apply it to Ho\v{r}ava--Witten theory though, one has to make sure that
the equations of motion allow a solution for which the function
$a_L$ goes to zero at the boundaries or else set $a_L = 0$
everywhere.

Let us also note that the $\overline{\Psi}{}^P\Gamma^{QR}\Psi^S G_{PQRS}$ term in the second line of 
(\ref{Supergravity}) can also contribute to a four dimensional mass term, because although 
the background value of the external flux is zero, it can
still have fluctuating components. The effect of this term, together with the Chern--Simons one $ C\wedge G\wedge G$, on the flux superpotential has been considered in \cite{HM}, 
for the case of $G_2$ structure. They lead to complexification of the geometric moduli.
We expect the same in our case too, but since we only focus on the geometric moduli here,
we will not go into details about the moduli of the $C$-field.

\subsubsection{Including warp factors}

Let us now turn to including non-trivial warp factors in the metric ansatz.
More precisely, we consider a metric of the form: \be \label{metans}
ds^2_{11} = e^{2b(x^m,x^{11})} \eta_{\mu \nu} dx^{\mu} dx^{\nu} +
e^{2f(x^m,x^{11})} g_{ln}(x^m, x^{11}) dx^l dx^n +
e^{2k(x^m,x^{11})} (dx^{11})^2 \, , \ee where $\mu$, $\nu$ still run
over the four external dimensions, whereas the internal indices are
split into six-dimensional ones ($m$, $n$,...) and $x^{11}$. The
six-dimensional space with metric $g_{ln}$ admits $SU(3)$ structure
and we denote the forms that the latter preserves by $J$ and
$\Omega$. In terms of these and $dx^{11}$, we can express the
seven--dimensional $SU(3)$ invariants $\tJ$, $\tO$ and $v$ as: 
\be
\label{resc} \tilde{J} = e^{2f} J \, , \qquad \tilde{\Omega} =
e^{3f} \Omega \, , \qquad v = e^k dx^{11} \, . 
\ee 
These rescalings can be seen as arising from the decomposition of the 11--dimensional
$\Gamma_M$ in the presence of warp factors (see appendix \ref{Relations}), or more
simply as induced by appropriate Weyl rescalings of the direct product ansatz 
(\ref{directprod}). As in the direct product case, the canonical normalization of
the four--dimensional gravitino kinetic term leads to a constraint on the coefficients
$a_L,a_R$ in the gravitino ansatz (note that we still normalize $\theta^\dagger\theta=1$), which  is:
\be \label{warpnormalisation}
|a_L|^2+|a_R|^2=e^b \, .
\ee
To obtain the full superpotential, we have to supplement the substitution of
(\ref{resc}) in (\ref{Wfl}) and (\ref{Wgeom}) with the additional
contribution coming from the modification of the derivative, $D_M
\rightarrow D_M^{(warp)}$, in the 11d kinetic term (\ref{kin}) which
is due to the presence of the warp factors. To extract the
coefficient of the four-dimensional gravitino mass term, we only
need the components along the internal directions: 
\be
\begin{split}
D_m^{(warp)} &= D_m + \frac{1}{2} \Gamma_m{}^n \partial_n f 
+ \frac{1}{2} \Gamma_m{}^{11} \partial_{11} f = 
D_m + \frac{1}{2} \Gamma_m{}^a \partial_a f \, ,  \\
D_{11}^{(warp)} &= \partial_{11} + \frac{1}{2} \Gamma_{11}{}^n
\partial_n k = D_{11} + \frac{1}{2} \Gamma_{11}{}^a \partial_a k \,
, 
\end{split}
\ee 
where, as before, $a = (\{n\},11)$. Therefore, we obtain for
the contribution of the extra terms in $D_a^{(warp)}$ (compared to
$D_a$) to the superpotential density the following: 
\be 
w^{(warp)} =
\frac{1}{2} a_R a_L^* \left[(\theta^{\dagger} \gamma^l \gamma_l{}^a
\theta + \theta^T \gamma^l \gamma_l{}^a \theta^*) \partial_a f +
(\theta^{\dagger} \gamma^{11} \gamma_{11}{}^a \theta + \theta^T
\gamma^{11} \gamma_{11}{}^a \theta^*) \partial_a k\right] = 0\, . 
\ee
The fact that this expression vanishes is easy to see by considering
$\theta^\dagger\gamma^a\theta+\theta^T\gamma^a\theta^*=v^a-v^a=0$.
Hence, although it might have seemed that the warp factors lead to a
contribution of the form $(const_1 \,df + const_2 \,dk)\wedge \tJ
\wedge \tJ \wedge \tJ$, this does not happen; as in \cite{DG}, they
do not affect the form of the superpotential.

To recapitulate, the superpotential generated in an $\Ncal=1$ flux
compactification of M--theory on an internal space with $SU(3)$
structure and metric of the form (\ref{metans}) is given by the following 
expression:
\bea \label{finalW} 
W &=&W^{(flux)} + W^{(geom)} = \nn \\ 
&=& \frac{1}{4} \int \left[ (a_R)^2 \,\,G\wedge \tO 
+ (a_L^*)^2 \,\,G \wedge \btOmega \right] 
+ \frac{i}{4} \int \left[ (a_R)^2 \,v \wedge d\tJ \wedge \tO 
+ (a_L^*)^2 \,v \wedge d\tJ \wedge \btOmega \right] \nn \\ 
&+& \frac{1}{2} \int a_L^* a_R \,\,G \wedge \tJ \wedge v 
- \frac{i}{16} \int a_L^* a_R [d\tO \wedge \btOmega 
+ d\btOmega \wedge \tO + 4 dv\wedge v \wedge \tJ \wedge \tJ] \nn \\ 
&-& \frac{1}{6} \int (a_L^* d a_R - a_R d a_L^*) \wedge \tJ \wedge \tJ \wedge
\tJ \, ,
\eea
where relations (\ref{resc}) should be substituted, and (\ref{warpnormalisation})
imposed. In (\ref{finalW}) we have collected all the terms that depend either
only on $a_L^*$ or only on $a_R$ in the first line. As already explained above, when specializing to 
Ho\v{r}ava--Witten theory, they are the only ones that contribute in the weakly coupled limit, 
in which either $a_L^*$ or $a_R$ is zero.

\subsubsection{Comparison with $G_2$ structure}

We end the current section with a comparison of our result to
the superpotential obtained from compactifications on $G_2$ structure manifolds. 
The latter was derived from dimensional reduction in \cite{HM}. 
Ignoring terms that come from fluctuations of the C-field (those  originating from the 11d 
Chern-Simons and $\overline{\Psi}{}^P\Gamma^{QR}\Psi^S G_{PQRS}$ terms, that were alluded to above), 
the result of \cite{HM} becomes:
\be \label{Sup}
W = \frac{1}{4} \left( \int G\wedge \Phi - \frac{i}{2} \int d\Phi \wedge \Phi \right) \, ,
\ee
where $\Phi$ is the $G_2$ structure defining 3-form.\footnote{Actually, in (\ref{Sup}) there is a sign difference compared to \cite{HM}, which is due to a sign difference between \cite{HM} and \cite{BCL} in the convention for the definition of $\Phi$ in terms of a spinor bilinear.} As anticipated, our superpotential
reduces to (\ref{Sup}) in the limit where $a_L=a_R^*$. However, the extra generality
of our ansatz is necessary to describe {\it heterotic} M--theory (rather than M--theory) 
compactifications. To see this, recall that any orientable seven--dimensional manifold 
admits both $G_2$ and $SU(3)$ structures and the relation between them is well-known:
\be \label{emb}
\Phi = {\rm Re} \,\tilde{\Omega} + \tilde{J}\wedge v \, .
\ee
Hence, one might have thought that the superpotential of interest for us could be 
obtained by substituting (\ref{emb}) in (\ref{Sup}). It is clear though, that in such 
a case one cannot recover the known weakly coupled limit. In other words, introducing 
different functions $a_L \neq a_R$ in the spinor ansatz is of crucial importance for 
the eleven--dimensional compactification to describe the strongly coupled limit of the
$E_8 \times E_8$ heterotic string.

\section{On moduli stabilization}
\setcounter{equation}{0}

In this section we would like to study the minima of the 4d $\Ncal=1$
supergravity potential determined by $W$ and $K$ of the previous
section in the standard way:\footnote{We define, as usual,  $K_A=\p_A K$, $K_{A\bar{B}}=\p_A\p_{\bar{B}}K$ 
and $D_A=\p_A+K_A$.} 
\be \label{U}
U = e^{K} (K^{A\bar{B}} D_A W D_{\bar{B}} \bar{W} - 3 |W|^2)\;,
\ee 
and the issue of moduli stabilization for these backgrounds.
However, in general the moduli spaces of $SU(3)$ structure
manifolds are not well-understood. In order to make progress, one
can consider approximate solutions consisting of turning on small
fluxes and neglecting their backreaction on the geometry.\footnote{This is the approach of \cite{DGKT}, for example.} Then the
light fields in the low energy effective theory are described by the
K\"{a}hler and complex structure moduli of the Calabi--Yau and the orbifold
length. The latter constitute the light field content of the
four-dimensional theory obtained from compactifications in which the
flux--induced background deformation of the zeroth order internal
space $CY(3)\times S^1/\bb{Z}_2$ is entirely encoded in warp
factors. This is the case with the Curio-Krause solution \cite{CK} (see also 
\cite{CK2} for more details on this background) that was employed to argue for the existence of dS vacua \cite{BCK} and assisted inflation \cite{BBK} in heterotic M--theory.

We start by analysing this background. More precisely, we will turn on a supersymmetry breaking flux on top of it and study whether the orbifold length can be stabilized at a dS vacuum. (This is not trivial as broken supersymmetry does not necessarily imply a de Sitter minimum of the scalar potential.) We also comment on the effect for moduli stabilization of the $R^4$ term of eleven--dimensional supergravity, which was shown in \cite{AV} to still be compatible with only warp-factor deformations. In addition, we show in passing how it corrects the generalized Hitchin flow equations to ${\cal O}(\kappa^{4/3})$. Finally, we comment on the general case of non--K\"{a}hler deformations of the initial $CY(3)$. 

\subsection{Warp-factor deformations} \label{Warp}

Since now we are interested in backgrounds that differ from the zeroth
order one only by warp factors, we take $g_{ln}$ in
(\ref{metans}) to be a $CY(3)$ metric which is independent of
$x^{11}$. As usual, one can introduce the K\"{a}hler ($t_i$) and complex structure ($z_{j}$) moduli of the $CY(3)$ via the decompositions: 
\be \label{mod}
J = \sum_{i=1}^{h^{1,1}} t_i w_i \, , \qquad \Omega = \sum_{j=1}^{h^{2,1}+1} (z_{j} \alpha_{j} - g_{j} \beta_{j}) \, , 
\ee 
where $\{w_i \}$ is a basis for $H^{1,1}$ and $\{ \alpha_{j}, \beta_{j} \}$ a basis for $H^3$. However, via the warp factors $f$ and $k$ in (\ref{resc}) the seven--dimensional forms $\tJ$ and $\tO$ do depend on $x^{11}$. So in order to obtain the moduli of the effective four--dimensional theory one has to perform a proper averaging over $x^{11}$. Recall that the coefficients $g_j$ in (\ref{mod}) are not independent coordinates on the complex structure moduli space as they can be expressed in terms of $z_j$, i.e. $g_j= g_j(\{z_i\})$ \cite{CdO}. Also, among the $z_j$'s there is one extra variable since the dimension of the moduli space they parametrize is $h^{2,1}$ and not $h^{2,1}+1$. That extra variable is related to rescaling of $\Omega$ and hence to the volume modulus of the Calabi--Yau as $\Omega \wedge \bar{\Omega} \approx vol_{CY}$. 

In the following we will assume, as in \cite{BCK,BBK}, that the K\"{a}hler and complex structure moduli have already been fixed. Given that, we will  address the question whether the orbifold length can be stabilized at a dS minimum by the superpotential (\ref{finalW}) only, i.e. without taking into account non--perturbative effects. One could also study whether other (than the orbifold length) moduli can be fixed by this superpotential. For the backgrounds of interest in this section, such will be the case with the complex structure moduli, as we will see in a moment. In fact, their stabilization would go exactly as in the weakly coupled string case and so we do not elaborate more on that.

\subsubsection{Curio-Krause background}

In order to specialize the considerations of Section 2 to the all--orders (in $\kappa^{2/3}$) background of \cite{CK}, we set $a_L^* = 0$ and for convenience denote $a_R \equiv a$. Therefore, we are left with a
superpotential of the form: 
\be \label{W_CK}
W = \frac{1}{4} \int e^b [ G\wedge e^{3f}\Omega + i e^k dx^{11}\wedge d\tJ \wedge e^{3f}\Omega] \, , 
\ee 
where we have substituted the relation $a = e^{b/2}$ from the previous section. The same relation follows from supersymmetry for the background of \cite{CK}. Let us introduce the following notation for the general decomposition of the four--form field strength $G$ w.r.t. the vector $v$: 
\be 
G (x^m, x^{11}) = G'(x^m,x^{11}) + v \wedge H(x^m,x^{11}) \, , 
\ee 
where $i_v G' =0$ and $i_v H = 0$. In the supersymmetric strong--coupling solution of \cite{CK}, $G$ does not have a leg in the $x^{11}$ direction (i.e. $H = 0$) and $G' = G' (x^m)$. Also, all warp factors are independent of $x^m$ 
(so, in particular, $\diff \tJ=2(\p_{11}f) e^{2f}\diff x^{11}\wedge J$ which vanishes
in (\ref{W_CK})) and
\be \label{CKb}
k(x^{11}) = -b(x^{11}) \, . 
\ee
Clearly then, the superpotential is zero as expected for a supersymmetric Minkowski vacuum.

Let us now turn on small supersymmetry breaking flux. In particular, since a $(0,3)$ flux component breaks supersymmetry \cite{DP}, we choose $H$ to be of the form
\be
H = p \,\btOmega \, ,
\ee 
where $p$ is an unspecified parameter.
Since, following \cite{DGKT}, we neglect the backreaction on the geometry we still have the background of \cite{CK}. Therefore, the second term in (\ref{W_CK}) vanishes again (together with its derivatives). However, from the first one we obtain:
\be \label{W_V}
W = \frac{p}{4} \int e^{k+b} dx^{11} \int e^{6f} \bar{\Omega} \wedge \Omega 
= 2\, i\, p \int dx^{11} V_{(6)}(x^{11}) = 2 \,i \,p \,{\cal L} \,{\cal V} ({\cal L}, {\cal V}_v) \, .
\ee
(We have used the standard six--dimensional normalization 
$\int e^{6f}\bar{\Omega}\wedge\Omega=8i V_{(6)}(x^{11})$, which is consistent with (\ref{volume}).) 
The function ${\cal V} ({\cal L}, {\cal V}_v)$ is known to be \cite{BCK}:
\be \label{V_BCK}
{\cal V} ({\cal L}, {\cal V}_v) = {\cal V}_v \left(1 - {\cal S} {\cal L} + \frac{1}{3} ({\cal S} {\cal L})^2\right) \, ,
\ee
where ${\cal V}_v$ is the Calabi--Yau volume at the visible boundary and ${\cal S}$ is a flux parameter that is independent of $x^{11}$. The variables ${\cal V}$ and ${\cal L}$ are dimensionless and are obtained by conveniently measuring the CY volume $V_{(6)}$ and orbifold length $L$ in terms of the following dimensionful reference quantities \cite{MPS}:
\be
v = 8 \pi^5 l_{11}^6 \, , \qquad l = 2 \pi^{1/3} l_{11} \, ,
\ee
where $l_{11}$ is the eleven--dimensional Planck length (e.g. ${\cal L} = L/l$, where $x^{11}\in [0,L]$).

Before going on to study the minimization of the scalar potential $U$ w.r.t. the orbifold length ${\cal L}$, two remarks are in order. First, the universal moduli $S$, $T$ of the zeroth order background $CY(3)\times S^1/\bb{Z}_2$ are now functions of the moduli ${\cal V}_v$ and ${\cal L}$ via
\be
{\rm Re} S = {\cal V} ({\cal L}, {\cal V}_v) \, , \qquad {\rm Re} T = {\cal L} \,{\cal V}^{1/3} ({\cal L}, {\cal V}_v) \, .
\ee
Since we assume that ${\cal V}_v$ has already been stabilized,\footnote{This issue was studied in great detail for the weakly coupled heterotic string in \cite{BBDG,BBDP}.} from now on everything is a function of a single variable ${\cal L}$. And second, from (\ref{W_V}) it is immediately obvious that the decompactification limit ${\cal L} \rightarrow \infty$, in which one expects to find a global supersymmetric minimum as in \cite{BCK}, is also a limit in which the susy breaking flux that we have turned on becomes very large. So its backreaction on the geometry cannot be neglected anymore and hence the approximation in which we are working (i.e. the Curio--Krause background plus additional small flux) breaks down.

Now let us turn to the investigation of the scalar potential $U$. In principle, the indices $A$, $\bar{B}$ run over all moduli, which in our case also includes the complex structure ones. However, we have assumed that the latter are stabilized at a supersymmetric point and so $D_i W = 0$ where $i = 1,...,h^{2,1}$. Therefore, using
\be \label{Kpot}
K = K_0 = -\ln (S+\bar{S}) - 3 \ln (T + \bar{T})
\ee
and (\ref{W_V}), (\ref{V_BCK}) we find\footnote{Notice that the axionic scalars, which make up the imaginary parts of $S$ and $T$, appear neither in $K$ nor in $W$ in the case under consideration.}
\be \label{Upot}
\begin{split}
U &= e^K\left[K^{S\bar{S}}D_SWD_{\bar{S}}\bar{W}+K^{T\bar{T}}D_TWD_{\bar{T}}\bar{W} - 3 |W|^2 \right]\\
&=\frac{p^2}{4 {\cal V}^2 {\cal L}^3} \left[{\cal L}^2 {\cal V}^2 + 4 {\cal V}^2 \left(\frac{{\cal L} {\cal V}}{{\cal V}'} + \frac{{\cal V}^2}{{\cal V}'^2}\right) + \frac{4}{3} {\cal L}^2 {\cal V}^{2/3} W_T^2 - 4 {\cal L}^2 {\cal V}^{4/3} W_T\right] \, ,
\end{split}
\ee
where
\be
W_T \equiv \frac{{\cal L} {\cal V}' + {\cal V}}{{\cal V}^{1/3} + \frac{1}{3} {\cal L} {\cal V}^{-2/3} {\cal V}'}
\ee
and ${\cal V}' \equiv \partial {\cal V}/ \partial {\cal L}$. The equation
\be
\frac{\partial U}{\partial {\cal L}} = 0
\ee
is quite messy, despite being purely algebraic. In particular, its numerator is a polynomial of 11th degree.\footnote{Its coefficients are not at all illuminating and so we opt not to write them down.} Hence it is not possible to write an analytic expression for its solutions.  Nevertheless, one can find them numerically. It turns out that from the eleven roots three are real and positive and the rest are pairs of complex conjugates. Since we do not want ${\cal L}$ to get too big in order to be within the range of validity of our approximation, we are left with only the smallest of the real roots, that we denote by ${\cal L}_0$, as a possible solution. Its value depends on the parameter ${\cal S}$. Realistic (i.e. phenomenologically preferred) values of ${\cal S}$ are between 1/30 and 2 or so \cite{BCK}.\footnote{The dependence on the value of ${\cal V}_v$ is only implicit through ${\cal S}$ \cite{BCK}. The explicit dependence on ${\cal V}_v$ cancels out of (\ref{Upot}).} We have investigated the values of the potential and its second derivative for various values of ${\cal S}$ in the above range and generically find that:
\be
U({\cal L})|_{{\cal L}={\cal L}_0} > 0 \qquad {\rm and} \qquad \frac{\partial^2 U}{\partial {\cal L}^2}|_{{\cal L}={\cal L}_0} > 0 \, .
\ee
Two examples are:
\be
{\cal S} = \frac{1}{11} \qquad \Rightarrow \qquad {\cal L}_0 = 12.75 \, , \,\,\,\, 
U({\cal L}_0) = 1.84\times10^{-2} p^2 \, , \,\,\,\, 
U''({\cal L}_0) = 7.36\times10^{-2} p^2
\ee
and
\be
{\cal S} = \frac{1}{3} \qquad \Rightarrow \qquad {\cal L}_0 = 3.48 \, , \,\,\,\, 
U({\cal L}_0) = 6.74\times10^{-2} p^2 \, , \,\,\,\, 
U''({\cal L}_0) = 0.42 p^2 \, .
\ee
Hence it looks like we have found a local de Sitter minimum.

\begin{figure}[t]
\begin{center}
\raisebox{0.28\textheight}{$\frac{U(\cal L)}{p^2}\!$}
\includegraphics[width=0.40\textwidth,height=0.3\textheight]{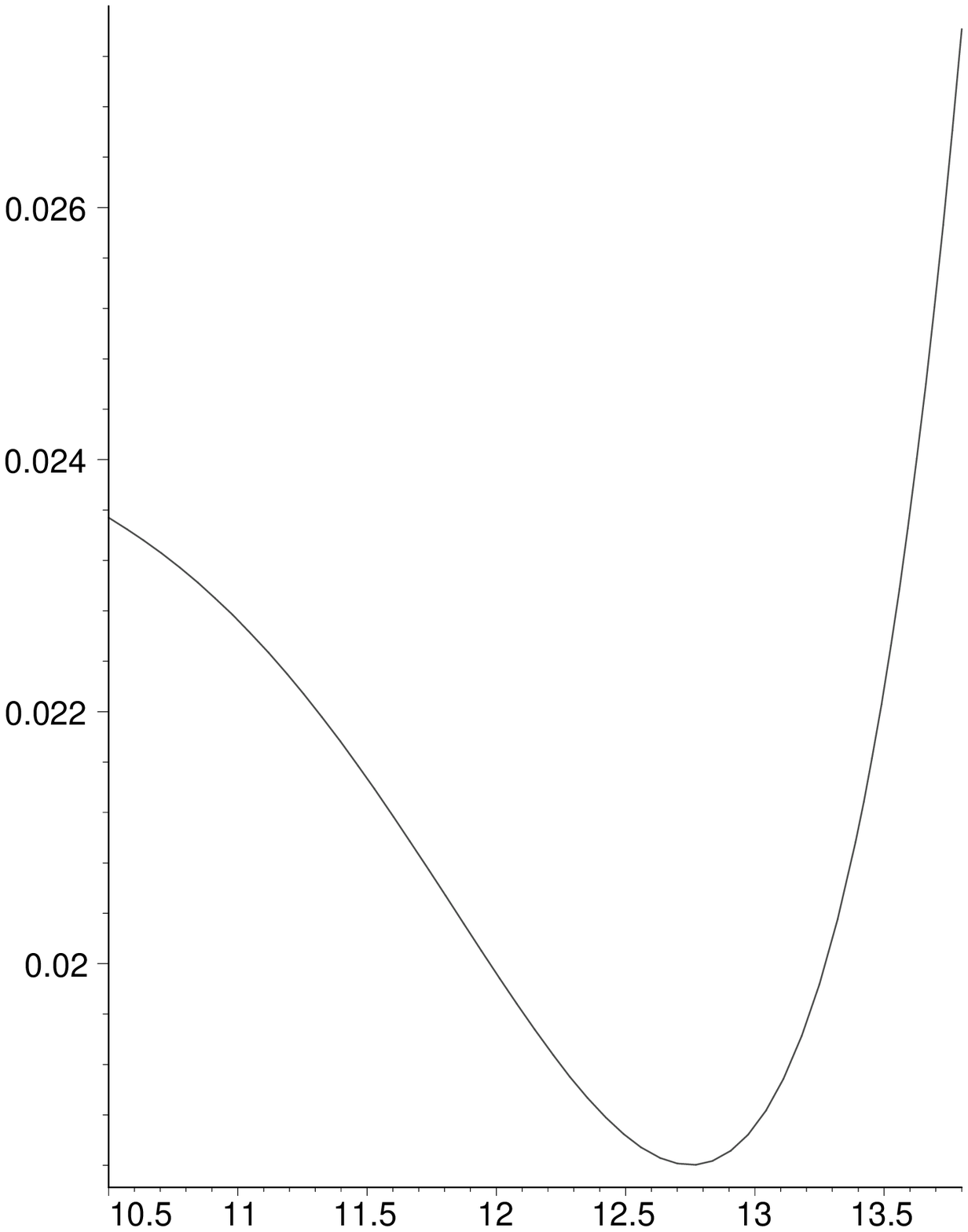}
\raisebox{-10pt}{$\!\!\!\!\!\!\cal L$}\hspace{.6cm}
\raisebox{0.28\textheight}{$\frac{U(\cal L)}{p^2}\!$}
\includegraphics[width=0.40\textwidth,height=0.3\textheight]{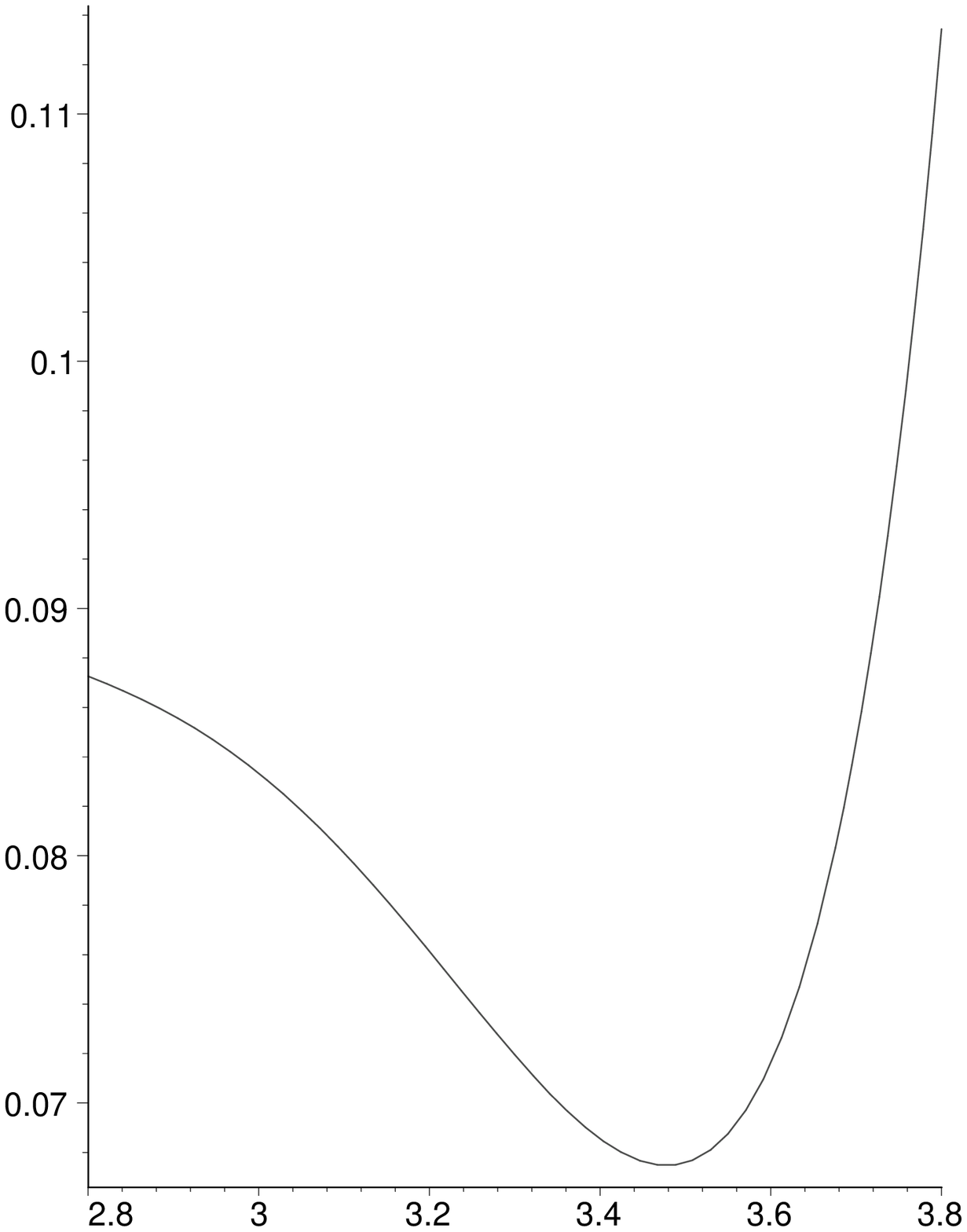}
\raisebox{-10pt}{$\!\!\!\!\!\!\cal L$}
\end{center}
\caption{The potential $U(\mathcal{L})$ (in units of $p^2$) as a function of the orbifold length $\cal L$.
The graph on the left is calculated for a value $\mathcal S=1/11$, while that on the 
right for $\mathcal S=1/3$.}
\end{figure}

Unfortunately though, things are not that simple. Namely, the Curio--Krause background develops a curvature singularity when $x^{11} = 1/{\cal S}$ \cite{CK, BCK}. Therefore, there is an upper bound on the physically allowed values of the orbifold length: ${\cal L} < {\cal L}_{max} = 1/{\cal S}$. The dS minimum that we obtain always occurs at values ${\cal L}_0$ which are larger than ${\cal L}_{max}$, although ${\cal L}_0$ does shift towards ${\cal L}_{max}$ as one increases ${\cal S}$. The conclusion is then that the flux superpotential alone is not enough to stabilize the orbifold length at a dS vacuum within the allowed range. However, the behaviour of the function $U({\cal L})$, see Figure 1, suggests that one may be able to lower ${\cal L}_0$ enough by including gaugino condensation (whereas the other crucial non--perturbative effect in \cite{BCK}, open membrane instantons, seems to be washed out by the contribution of the flux superpotential.). We defer more detailed discussion on that to Section 5.

\subsubsection{$R^4$ term and torsion classes}

Including corrections to the Curio--Krause background, that are due to the $R^4$ term of eleven--dimensional supergravity, still results in a purely warp--factor deformation of the zeroth order internal space $CY(3)\times S^1/\bb{Z}_2$ \cite{AV}. In fact, the solution in this case is only known to ${\cal O}(\kappa^{4/3})$ \cite{AV}. Whether or not it can be extended to higher orders is beyond the scope of the present paper.\footnote{Such an extension has to be done order by order, taking into account other relevant higher derivative terms like \cite{GKV}.} Nevertheless, it can be useful in building intuition about supersymmetric backgrounds with both $G' \neq 0$ and $H \neq 0$ since it can accommodate both flux components turned on. In particular, one could perform in this background the same  computation as in Subsection 3.1.1 in search for dS vacua. In doing so, one also has to take into account the correction to the K\"{a}hler potential (\ref{Kpot}), that originates from the $R^4$ term in the effective action \cite{AV}.\footnote{It may seem that this correction can be absorbed by a $\kappa$-dependent field redefinition of the universal  modulus $S$, but this is deceptive. The reason is that the gauge kinetic functions also depend on $S$. As a result, one cannot redefine away the $R^4$-induced correction but only shift it from the K\"{a}hler metric into the gauge kinetic functions. This, however, would spoil the correspondence with the weakly coupled heterotic string. So the natural place of the $R^4$-induced correction is indeed in the K\"{a}hler potential.} However, we do not expect the result for the value of the orbifold length at the extremum of $U$, ${\cal L}_0$, to shift in the desired direction. The reason is that the $R^4$ contribution was shown in \cite{AV} to move ${\cal L}_0$ towards larger values in the case of a superpotential generated entirely by non--perturbative effects. It is still possible though that the $R^4$ term influences differently the behaviour of $W_{flux}$; this is certainly worth investigating and we hope to return to it in the future. Instead, here we want to address the role of the $R^4$ term in the $SU(3)$ structure description of the internal manifold. In particular, we want to find out how it modifies the torsion classes of the latter (to order $\kappa^{4/3}$).

The background of \cite{AV} still satisfies $a_L=0$. Hence, it belongs to the class for which \cite{DP} derived 
`generalized Hitchin flow' equations, albeit in the absence of the $R^4$ corrections. These  equations are relations between $dJ$, $d\Omega$, $\partial_{x^{11}} J$, $\partial_{x^{11}} \Omega$ and the various flux components and warp factors, that follow from the supersymmetry condition $\delta_M \Psi = 0$. In other words, they encode the information about what six-dimensional manifolds together with what fibrations of them along $x^{11}$ constitute solutions compatible with supersymmetry. However, the $R^4$ term in the 11d effective action does modify the supersymmetry variation of the gravitino. So it should affect the above relations, which in particular means the relation between the torsion classes of the six-dimensional manifold and the background flux components. This is in line with the arguments in Section 4 of \cite{CCDL} about the influence of higher derivative terms on the relation between torsion classes and flux components for the weakly coupled heterotic string. As noted there, analyzing in full generality the contribution of the higher derivative terms is quite a daunting task. However, we will see below that in the particular case of interest for us it is rather easy to compute to ${\cal O}(\kappa^{4/3})$ the $R^4$--induced corrections.

Despite the fact that the complete supersymmetry transformations in the presence of the $R^4$ term are still unknown, it was shown in \cite{CFPSS,LPST1,LPST2} that for string and M--theory compactifications on special holonomy manifolds (more precisely: $CY(3)$, $G_2$ and $Spin(7)$) the gravitino transformation is modified by the $R^4$ term in the following way:
\be \label{grvar}
\delta_M \Psi = (D_M + \varepsilon(\nabla^N R_{MKP_1P_2}) R_{NLP_3P_4} R^{KL}{}_{P_5P_6} \Gamma^{P_1...P_6})\eta \, ,
\ee
where $D_M$ is the usual covariant derivative extended with flux terms, which for 11d supergravity is
\be
D_M = \nabla_M + \frac{1}{288} (\Gamma_{MNKLP} - 8 \delta_{MN} \Gamma_{KLP}) G^{NKLP} \, ,
\ee
and $\varepsilon$ is a numerical constant times $\kappa^{4/3}$ in M--theory (and times $\alpha'^3$ in string theory). In the rest of this subsection we will work with accuracy to order $\kappa^{4/3}$. It is clear then that in the second term of (\ref{grvar}) one should substitute the zeroth order curvatures, which in our case are the curvatures for the direct product  background $M_4\times CY(3)\times S^1/\bb{Z}_2$. Hence (\ref{grvar}) simplifies to the following Killing spinor equation:\footnote{Recall that $m=1,...,6$ are the CY dimensions. Clearly, the term cubic in curvatures in (\ref{grvar}) does not contribute to any other directions.}
\be \label{Pm}
\delta \Psi_m = (D_m + P_m) \eta = 0\, ,
\ee
where
\be
P_m = -\frac{i}{2} \varepsilon J_{mn} \partial^n Q
\ee
with $Q$ being the Euler density in six dimensions and $J$ the K\"{a}hler form of the CY as before. 

It is very easy to trace how the new $P_m$ term in (\ref{Pm}) affects the considerations of \cite{DP}. Let us first summarize the calculations without $P_m$. From the supersymmetry condition $\delta \Psi_M = D_M \eta = 0$ and the expressions for the $SU(3)$ structure defining forms in terms of spinor bilinears one derives (3.16)-(3.18) of \cite{DP}: 
\bea \label{dforms}
dv &=& 2 v \wedge db\, , \nn \\
d\tJ &=& -4\tJ\wedge db - 2 *G\, , \nn \\
d\tO &=& 3 \tO \wedge db \, .
\eea
These equations can be decomposed into components along $v$ and orthogonal to $v$. The first set determines the fibration structure of the 6d manifold $Y$ along the interval (the so called `Hitchin flow'), whereas the second set -- the torsion classes of $Y$ in terms of flux components and warp factors. (We will refer to both sets as
 `generalized Hitchin flow' equations.) Let us now include the $P_m$ term. Recall, that the supersymmetry parameter $\eta$ is given in terms of the internal spinors $\theta$ and $\theta^*$ via the same kind of ansatz as the one for the four-dimensional gravitino. Taking into account $P_m$  clearly modifies the covariant derivative on $\theta$ in the following way: $\delta (\nabla_m \theta) = P_m \theta$. Since $P_m$ is purely imaginary, this implies that $\delta (\nabla_m \theta^{\dagger}) = - P_m \theta^{\dagger}$. Hence in the covariant derivative of any spinor bilinear of the form $\theta^{\dagger} \gamma_{m_1 ... m_p} \theta$ the total contribution of the $P_m$ term is zero. Due to (\ref{spbilin}), this means that $dv$ and $d\tJ$ do not get any correction. On the other hand, in the covariant derivative $\nabla_n$ of any bilinear of the form $\theta^T \gamma_{m_1 ... m_p} \theta$ one obtains the additional term $2P_n \theta^T \gamma_{m_1 ... m_p} \theta$. Therefore, the last equation in (\ref{dforms}) gets modified to
\be
d\tO = 3 \tO \wedge db + 2 P\wedge \tO \, ,
\ee
where we have introduced the notation $P \equiv P_m dx^m$. Since both $i_v P = 0$ and $i_v\tO=0$, the new term does not affect the fibration structure along $x^{11}$. However, it does contribute to the torsion classes of the six-dimensional slices. Comparing with (\ref{tor}), we see that it changes ${\cal W}_5$ as
\be
\delta {\cal W}_5 = - 2 P \, .
\ee

Let us again remind the reader that the above result is only valid to order $\kappa^{4/3}$. Most likely, at higher orders the 11d $R^4$ term will lead to additional corrections to the generalized Hitchin flow equations. Even in the simplest possible case, i.e. having warp factors only, given that generically the latter will depend on both $x^m$ and $x^{11}$ the cubic term in (\ref{grvar}) seems poised to modify both the fibration structure along $x^{11}$ and the torsion classes of the six-dimensional slices.

\subsection{General backgrounds}

In this subsection we comment on the superpotential generated by
more complicated backgrounds. An obvious generalization of the considerations of Subsection \ref{Warp} is to take the six-dimensional manifold with metric $g_{ln}$ to be non--K\"{a}hler. In this case the dependences on $x^{11}$ and $x^m$ do not factorize unlike in Subsection \ref{Warp}. Explicit solutions of this kind are not known at present. Hence, it is not possible to extract the explicit dependence of the superpotential on ${\cal L}$. However, it is clear that in principle $W$ does depend on it. Still keeping $a_L=0$, now both terms in 
\be 
\label{Wtwo} W = \int a^2 [ G\wedge \tO + \frac{i}{4} v \wedge d\tJ \wedge \tO] 
\ee 
are nonvanishing. Since all ingredients in (\ref{Wtwo}) depend on $x^{11}$, it seems in principle possible to stabilize the orbifold length. And due to the no-scale structure of the K\"{a}hler potential at leading order, the
scalar potential is positive definite. Whether or not it can have dS vacua,
apart from Minkowski ones, cannot be determined without a particular background at hand.

The most general case is clearly given by both $a_L, a_R \neq 0$. Now all terms in (\ref{finalW}) are non-vanishing. As we have already mentioned, the moduli spaces of non--K\"{a}hler manifolds are not well--understood at present. Nevertheless, using loosely the terminology appropriate for compactifications conformal to Calabi--Yau, we can see that the superpotential seems to depend both on the `complex structure' moduli and on the `K\"{a}hler' ones as in type IIA flux compactifications. Hence, it is conceivable that the
flux superpotential can stabilize all geometric moduli as in \cite{DGKT}, which, in particular, includes 
the orbifold length as
all quantities generically depend on it. Let us stress again that the M--theory superpotential (\ref{finalW}) in the case of both $a_L$ and $a_R$ non-vanishing is applicable to heterotic M--theory only when the appropriate
boundary conditions are allowed (including $a_L \rightarrow 0$ as
one approaches the visible boundary). 

Final remark in this section: In the conventions of \cite{LS}, the case $a_L, a_R \neq 0$ corresponds to different norms of the two internal singlet spinors. It was shown in \cite{LS} that this is a necessary but not sufficient condition
for supersymmetric AdS vacua. On the other hand, supersymmetric Minkowski vacua can exist for either one of $a_L, a_R$ vanishing or both of them nonzero. Furthermore, \cite{LS} showed that for the perturbative heterotic M--theory background of \cite{LOW}, in which the six-dimensional space fibered over the interval is a generic non--K\"{a}hler manifold, only one of $a_L$ and $a_R$ is nonvanishing; let us say for definiteness that $a_L = 0$. This background is the most general perturbative supersymmetric solution with four-dimensional Minkowski space to first order in the $\kappa^{2/3}$ expansion \cite{LS}. Since it is not known at higher orders though, clearly the considerations of \cite{LS} do not contradict the existence of heterotic M--theory solutions with both $a_L, a_R \neq 0$ as long as the first nontrivial contribution to $a_L$ starts at order $\kappa^{4/3}$ or higher.

\section{K\"{a}hler potential and charged matter vevs}
\setcounter{equation}{0}

In this section we discuss what the implications of the presence of nonzero flux--induced superpotential are on the role of perturbative corrections to the K\"{a}hler potential and on the stabilization of the charged matter fields in heterotic M--theory.

Let us start by recalling that in the effective action derived in \cite{LOW} the only contribution to the superpotential is given by
\be \label{Wtree}
W_{tree} = \Lambda_{IJK} C^I C^J C^K \, ,
\ee
where, up to a numerical coefficient, $\Lambda_{IJK}$ are the Yukawa couplings and $C^I$ are the charged matter superfields that arise from the boundary $E_8$ gauge multiplets. Clearly, $W_{tree}$ does not depend on the orbifold length ${\cal L}$ nor on any other K\"{a}hler or complex structure modulus. In \cite{BCK} the dependence on ${\cal L}$ was introduced via the non--perturbative contribution to the superpotential due to open membrane instantons ($W_{OM}$) and gaugino condensation ($W_{GC}$). 

One may wonder, though, whether including perturbative contributions to the potential $U$ in (\ref{U}), due to corrections to the K\"{a}hler potential,\footnote{Recall, that the superpotential is not expected to get perturbative corrections because of the axionic shift symmetries arising from the gauge invariance of the 11d three--form field $C$.} will not change qualitatively the behavior of $U$ as for the IIB string \cite{BB}. Indeed, similarly to the K\"{a}hler potential correction of \cite{BBHL}, that was so crucial in the IIB case, there is also a correction to the K\"{a}hler potential of the universal moduli of heterotic M--theory due to the 11d $R^4$ term \cite{AV}. Moreover, in \cite{CKM} it was argued that the general structure of this K\"{a}hler potential in heterotic M--theory is given by a series in powers of $1/{\rm Re}(S)$ and $1/{\rm Re}(T)$ with leading correction terms:
\bea
\delta K &=& \frac{A_{03}}{[4\pi^2 {\rm Re}(T)]^3} \left[ 1 + {\cal O} \left( \frac{1}{4\pi^2 {\rm Re}(T)} \right ) \right] \nn \\
&+& \frac{A_{11}}{4\pi^2 {\rm Re}(S)} \left[ 1 + {\cal O} \left( \frac{1}{4\pi^2 {\rm Re}(S)}, \frac{1}{4\pi^2 {\rm Re}(T)} \right ) \right] \nn \\
\delta Z &=& \frac{3}{T + \bar{T}} \frac{B_{03}}{[4\pi^2 {\rm Re}(T)]^3} \left[ 1 + {\cal O} \left( \frac{1}{4\pi^2 {\rm Re}(T)} \right ) \right] \nn \\
&+& \frac{3}{2} \frac{B_{10}}{{\rm Re}(S)} \left[ 1 + {\cal O} \left( \frac{1}{4\pi^2 {\rm Re}(S)}, \frac{1}{4\pi^2 {\rm Re}(T)} \right ) \right] \, , \label{delK}
\eea
where
\be
K = K_0 + \delta K + \left( \frac{3}{T + \bar{T}} + \delta Z \right) |C|^2
\ee
with
\be
K_0 = - \ln (S + \bar{S}) - 3 \ln (T + \bar{T}) \, .
\ee
The correction found in \cite{AV} contributes to the $A_{11}$ coefficient. $B_{10}$ was determined in \cite{LOW}. The coefficients $A_{03}$ and $B_{03}$ were also argued in \cite{CKM} to be non-vanishing.

However, as was shown in \cite{AV}, $\delta K(A_{11})$ affected the behavior of the scalar potential $U$ for the warp-factor-deformed background of \cite{CK} only by a small percentage. The same will be true also for any other term in (\ref{delK}). The simple reason is that since the charged fields $C^I$ were stabilized at a nonzero but {\it very} small value \cite{BCK}, one could completely ignore their contribution to $U$ when addressing the stabilization of ${\cal L}$. In particular, one could drop $W_{tree}$ in (\ref{Wtree}). Due to that, there was no other contribution to the superpotential than $W_{non-pert} = W_{OM} + W_{GC}$. So the corrections to the K\"{a}hler potential $\delta K$, which are already suppressed w.r.t. to $K_0$, get even more suppressed in $U$ as to linear order in $\delta K$ one finds schematically: $\delta U \sim \delta K |W_{non-pert}|^2$.

Clearly, if there is a nonzero contribution to the superpotential due to background flux and the related backreaction of the geometry ($W_{flux} \equiv W^{(flux)} + W^{(geom)}$), then the above argument is no longer valid. I.e. terms of the form $\delta K |W_{flux}|^2$ dominate the terms $\delta K |W_{non-pert}|^2$ and hence can lead to significant qualitative changes w.r.t. the results of \cite{BCK, BBK}. Even more, the no-scale structure (i.e. independence) of $U$ on ${\cal L}$ is already violated by $W_{flux}$ for generic backgrounds and so this is really the leading contribution. In fact, as we noted in the previous sections, $W_{flux}$ depends on all geometric moduli and this opens up the possibility of stabilizing (for generic background flux) all of them without any non--perturbative effects similarly to type IIA string theory \cite{DGKT}.

Finally, there is one more significant difference with the only-warp-factor background deformation considered in \cite{BCK}, that we should point out. Namely, the presence of $W_{flux} \neq 0$ affects the vevs, $C_0^I$, of the charged matter fields $C^I$. In \cite{BCK} it was shown that the contribution of the $C^I$'s to the potential at the extremum is proportional to $|C_0|^4 \sim |W_{non-pert}|^4$ and hence is strongly (exponentially) suppressed w.r.t. to the other terms in $U$ which behave as $|W_{non-pert}|^2$. On the other hand, repeating the same considerations now (i.e. with $W_{flux}\neq 0$), one finds that these vevs are governed by $C_0 \sim W_{flux}$. Hence they are not exponentially suppressed anymore compared to the rest of the terms. It is still possible that they may be of subleading order in the expansion in terms of $1/{\cal V}$ and $1/{\cal V}_{OM}$, but this has to be checked on a case by case basis; it does not seem clear generically whether or not one can neglect the $C^I$ contribution while minimizing the effective potential $U$ with respect to the remaining moduli, including the orbifold length.

\section{Conclusions and discussion}

We considered M--theory compactifications on seven--dimensional manifolds with $SU(3)$ structure. We derived the corresponding flux--induced superpotential for the most general ${\cal N}=1$ spinor ansatz, giving the embedding of the four--dimensional gravitino into the eleven--dimensional one. This result can be specialized to heterotic M--theory, upon imposing the appropriate boundary conditions. Essential for that is the extra generality of the $SU(3)$ spinor ansatz as compared to the $G_2$ structure one. However, we have not considered the $E_8$ gauge multiplets propagating on the boundaries of the Ho\v{r}ava--Witten set up. In the weakly coupled $E_8\times E_8$ heterotic string these gauge fields were shown to lead to an additional contribution to the superpotential \cite{BBDP}. It is undoubtedly of great interest to recover the corresponding result within the strongly coupled description. In this work we have concentrated only on the geometric moduli. However, ignoring the moduli of the $C$-field  obscures the holomorphic nature of the superpotential. It is certainly worth investigating in detail how the axionic moduli complexify the geometric ones and restore holomorphicity. 

Although generically the moduli spaces of $SU(3)$ structure manifolds are not well-understood, in the cases when the compactification spaces are conformal to Calabi--Yau the relevant moduli are essentially known. For a heterotic M--theory background \cite{CK} of this kind, more precisely differing from the initial $M_4\times CY(3)\times S^1/\bb{Z}_2$ only by warp factors, we investigated the implications of the flux--induced superpotential for the  stabilization of the orbifold length ${\cal L}$. Our interest was to find out whether ${\cal L}$ can be stabilized at a de Sitter vacuum without including non--perturbative effects. The strategy was to turn on small supersymmetry breaking flux, whose backreaction on the geometry one ignores (as in the considerations of \cite{DGKT} for type IIA moduli stabilization), and study the resulting scalar potential. This is similar in spirit to the KKLT scenario \cite{KKLT}, where one has a supersymmetric AdS minimum and introduces probe anti--D3 branes to lift it to a dS one. Since branes and fluxes are often interchangeable via geometric transitions, it is conceivable that our setup may be mapped to a background with anti-branes in some dual string description. This is worth pursuing, but is well beyond the scope of the present paper. 

We did find a dS minimum of the scalar potential. However, it occurs at a value ${\cal L}_0$ of the orbifold length which is slightly larger than the physically allowed maximal one, ${\cal L}_{max}$. Strictly speaking, that means that the flux--induced superpotential alone is not enough in the search for dS vacua. On the other hand, the positive--cosmological--constant minimum that we found did not have to be at ${\cal L}_0$ so close to ${\cal L}_{max}$. It could have happened that ${\cal L}_0$ is orders of magnitude larger than ${\cal L}_{max}$ instead of only several percent. We believe that their proximity is an indication that the minimum we found is not too far from a physical dS vacuum obtained by adding subleading quantum effects. The relevant non--perturbative contributions to the superpotential, from gaugino condensation and open membrane instantons, were studied in great detail in \cite{BCK}. It was shown there that at smaller ${\cal L}$ open membrane instantons dominate the scalar potential and lead to increasing energy density as ${\cal L}$ decreases. On the other hand, at larger ${\cal L}$ gaugino condensation is dominant and leads to increasing energy as ${\cal L}$ increases. The two effects are balancing each other at some intermediate point, giving there a de Sitter minimum. The behaviour of our flux--induced scalar potential $U$ is such that the energy decreases as one approaches ${\cal L}_{max}$ from below, the lowest value of $U$ (in the physically allowed region) being attained at ${\cal L}_{max}$ (see Figure 1). Therefore the smallest values of $U$ are in the range in which membrane instantons become negligible. However, in this range the contribution of gaugino condensation drives the potential up as ${\cal L}$ increases \cite{BCK}. Hence, we would expect that the flux superpotential and gaugino condensation balance each other, producing a dS vacuum for orbifold length that is very close to ${\cal L}_{max}$.\footnote{Recall, that this is also the most phenomenologically--preferred value of the orbifold length.} This picture seems very appealing and certainly merits a detailed investigation. We hope to report on that in the future.

Another issue we touched upon is including $R^4$ corrections. Finding the appropriate background to all orders is a daunting task. But to order $\kappa^{4/3}$ the geometric deformation of the initial $M_4\times CY(3) \times S^1/\bb{Z}_2$ is still encoded in warp factors only \cite{AV}. We showed that, to this order, the $R^4$ term contributes to the generalized Hitchin flow equations, that determine the supersymmetric background, by only changing the torsion class ${\cal W}_5$ of the six-dimensional slice orthogonal to the orbifold direction. It is a natural question to ask how this higher derivative term affects the stabilization of the orbifold length. It is also worthwhile to address other higher derivative corrections as well as attempt to build the effective heterotic M--theory action at higher orders in the $\kappa^{2/3}$ expansion with the methods of \cite{IM,IM2}.

Finally, one could try to explore the stabilization of other (than the orbifold length) moduli in specific cases. The M--theory superpotential that we have found seems to depend on all geometric moduli and hence holds great promise for generic background fluxes. It would also be interesting to extract from the supersymmetric minima of this superpotential the  classification of \cite{BCL} of the ${\cal N}=1$ supersymmetric backgrounds in M--theory in terms of relations between torsion classes and flux components.

\subsection*{Note Added}
After submitting the first version of this paper to the arXiv, we were informed of the work 
in progress \cite{CLu} where the superpotential for M--theory on $SU(3)$ structure manifolds has 
also been derived (both for $\Ncal=1$ and $\Ncal=2$ supersymmetry in 4d), and is found to agree with 
our results in Section \ref{Superpotential}. We thank M. Cveti\v{c} for providing us with a preliminary draft of this work.

\section*{Acknowledgements}
We are indebted to G. Dall'Agata for an illuminating discussion. We would also like to thank K. Behrndt for very useful correspondence and J. Ward for conversations. The work of L.A. is supported by the EC Marie Curie Research Training Network MRTN-CT-2004-512194 {\it Superstrings}. K.Z. is supported by a PPARC grant ``Gauge theory, String Theory and Twistor Space Techniques''.

\appendix

\section{Conventions and definitions}
\setcounter{equation}{0}

 We use the notation $M,N,\ldots$ (ranging from 1 to 11) for eleven--dimensional indices,
$\mu,\nu,\ldots$ for four--dimensional ones, and $a,b,\ldots$ for seven--dimensional ones.
For six--dimensional indices (i.e. not including $x^{11}$), we write $m,n,p,q$ etc..

 We take the 11--dimensional gamma matrices to satisfy $\{\Gamma_M,\Gamma_N\}=2g_{MN}$ and to
be real. When reducing on a direct product metric, we decompose them as
\be
\Gamma_\mu=\gamma_\mu\otimes{1\!\!1}, \quad \text{and}\quad \Gamma_a=\gamma\otimes \gamma_a \,.
\ee
However, when reducing on a warp factor metric of the type
\be
\diff s^2=e^{2b}\diff s_{(4)}^2+e^{2f}\diff s_{(6)}^2+e^{2k}(\diff x^{11})^2 \, ,
\ee
in order to retain $\{\gamma_\mu,\gamma_\nu\}=2g_{\mu\nu}^{(4)}$ etc., we need to 
decompose $\Gamma_M$ as
\be
\Gamma_\mu=e^{b}\gamma_\mu\otimes{ 1\!\!1},\quad
\Gamma_a=\gamma\otimes e^f\gamma_a \quad
\text{and} \quad\Gamma_{11}=\gamma\otimes e^k\gamma_{11} \, .
\ee
This can also be seen by first reducing on a direct product metric and then performing
a rescaling of the metric to introduce the warp factors. 
(Note that $\gamma=\sqrt{-g_{(4)}}\epsilon_{\mu\nu\rho\sigma}
\gamma^{\mu}\gamma^{\nu}\gamma^{\rho}\gamma^{\sigma}$ is invariant under taking 
$\diff s_{(4)}^2\rightarrow e^{2b}\diff s_{(4)}^2$).
The following identities involving the $\Gamma_M$ are frequently useful 
(we define $\Gamma_{M_1\cdots M_p}\equiv\Gamma_{[M_1}\Gamma_{M_2}\cdots\Gamma_{M_p]}=
1/p![\Gamma_{M_1}\Gamma_{M_2}\cdots\Gamma_{M_p}\pm\ldots]$):
\be \begin{split}
\Gamma_M\Gamma_{N_1\cdots N_p}&=\Gamma_{MN_1\cdots N_p}+p \,g_{M[N_1}\Gamma_{N_2\cdots N_p]}\;, \\
\Gamma_{N_1\cdots N_p}\Gamma_M&=\Gamma_{N_1\cdots N_pM}+p \,\Gamma_{[N_1\cdots N_{p-1}} g_{N_p]M}\;. \end{split}
\ee
In 7 dimensions the gamma matrices are taken to be imaginary and antisymmetric
(and thus hermitian). Their antisymmetric products 
$\gamma_{a_1\cdots a_n}=\gamma_{[a_1}\gamma_{a_2}\cdots\gamma_{a_n]}$
satisfy the relation:
\be
(\gamma_{a_1\cdots a_n})^T=(-)^{\frac{n(n+1)}{2}}\gamma_{a_1\cdots a_n}
\ee
Seven--dimensional spinor products can be rearranged via the following 
Fierz identity: 
\be \label{Fierz}
\begin{split}
\theta_1^\dagger\Mcal\theta_2\theta_3^\dagger\Ncal\theta_4=
\frac{1}{8}
&\left(\theta_1^\dagger\theta_4\theta_3^{\dagger}\Ncal\Mcal\theta_2
+\theta_1^\dagger\gamma_a\theta_4\theta_3^{\dagger}\Ncal\gamma^a\Mcal\theta_2\right.\\
&\left.-\half\theta_1^\dagger\gamma_{ab}\theta_4\theta_3^{\dagger}\Ncal\gamma^{ab}\Mcal\theta_2\right.
\left.-\frac{1}{6}\theta_1^\dagger\gamma_{abc}\theta_4\theta_3^{\dagger}\Ncal\gamma^{abc}\Mcal\theta_2
\right)\;,
\end{split}
\ee
where $\Mcal,\Ncal$ are arbitrary products of gamma matrices. Fierz rearrangements 
will be very useful when checking the properties of the spinor bilinears
defined in (\ref{spbilin}). 

The volume of the internal seven manifold is defined as
\be \label{volume}
V_{(7)}=\int *1 =\frac{1}{6}\int\tJ\wedge\tJ\wedge\tJ\wedge v
=\frac{i}{8}\int\tOmega\wedge\btOmega\wedge v
\ee
(all Hodge stars in the appendix will be seven dimensional). 
Given these normalizations, and the result (given below) that $\tJ^{ab}\tJ_{ab}=6$, 
we can find the duals of $\tJ$ and $\tJ \wedge \tJ$:\footnote{Recall that $\int \alpha_p\wedge (\ast \beta_p)
=\frac{1}{p!}\int\alpha_{a_1\cdots a_p}\beta^{a_1\cdots a_p} \ast 1$.}  
\be
*\tJ=\half\tJ\wedge\tJ\wedge v, \quad *(\tJ\wedge\tJ)=2\tJ\wedge v\;.
\ee
Also, since $\tOmega^{abc}\btOmega_{abc}=48$ (as follows from (\ref{Omn})), we find:
\be
*\tOmega=-i\tOmega\wedge v,\quad *\overline{\tOmega}=i\overline{\tOmega}\wedge v \, .
\ee
It is also easy to check, using $v^av_a=1$, that 
$*v=\frac{1}{6}\tJ\wedge\tJ\wedge\tJ$.

\section{Some useful formulae} \label{Relations}
\setcounter{equation}{0}

In this Appendix we expand on some results that are required for the calculations 
in section \ref{Superpotential}.
One can easily derive the very useful relations 
\be \label{dbilapp}
(dv)_{ab} = 2 \tau_{[ab]}{}^c v_c \, , 
\quad (d\tJ)_{abc} = 6 \tau_{[ab}{}^d \tJ_{|d|c]} \, , 
\quad (d\tOmega)_{abcd} = 12 \tau_{[ab}{}^e \tOmega_{|e|cd]} \, ,
\ee
by using  $SU(3)$ invariance, $\nabla^{(T)}\omega=0$, where $\nabla^{(T)}$ is the
connection with torsion and $\omega=v,\tJ$ or $\tOmega$. 
Alternatively one can also arrive at (\ref{dbilapp})  from the definition of the forms as
spinor bilinears. For instance, to show the second relation in (\ref{dbilapp}), we can use the definition 
of $\tJ$ from (\ref{spbilin}), and that 
$\nabla_a\theta=\frac{1}{4}\tau_{abc}\gamma^{bc}\theta$, along with
$\nabla_a\theta^\dagger=-\frac{1}{4}\tau_{abc}\theta^\dagger\gamma^{bc}$
(since $(\gamma_{bc})^\dagger=-\gamma_{bc}$) to compute:  
\be
\nabla_a\tJ_{bc}=\nabla_a(-i\theta^\dagger\gamma_{bc}\theta)=
\frac{i}{4}{\tau_{a}}^{de}\theta^\dagger(\gamma_{de}\gamma_{bc}-\gamma_{bc}\gamma_{de})\theta \, .
\ee
It is straightforward to show that
\be
\gamma_{de}\gamma_{bc}-\gamma_{bc}\gamma_{de}=2(g_{db}\gamma_{ce}+g_{dc}\gamma_{eb}
-g_{eb}\gamma_{cd}-g_{ec}\gamma_{db}) \, .
\ee
Using this and the antisymmetry of $\tau_{abc}$ in its
last two indices, and antisymmetrizing w.r.t. $a,b,c$ \,, we find:
\be
(d\tJ)_{abc}=3\nabla_{[a}\tJ_{bc]}
=\frac{3i}{4}{\tau_{[a}}^{de}(8g_{b|d|}\theta^\dagger\gamma_{c]e})
=6i {\tau_{[a}}^{de}g_{b|d|}(i\tJ_{c]e})
=-6{\tau_{[ab}}^{d}\tJ_{c]d} \, .
\ee

Using (\ref{Fierz}) it is straightforward to derive the following identities
for the $SU(3)$-defining forms:
\be
\tJ^{ab}\tJ_{bc}=-\delta^{a}_{\;\;c}+v^av_c
\ee
and
\be \label{Omn}
\tOmega^{abc}\overline{\tOmega}_{abd}=-8i\tJ^c_{\;\;d}+8\delta^c_{\;\;d}-8v^cv_d \, .
\ee
These formulae imply, upon using the normalization $v^av_a=1$, that $\tJ^{ab}\tJ_{ab}=6$ and $\tOmega^{abc}\btOmega_{abc}=48$, which coincide with  the standard normalizations for \emph{six}--dimensional $SU(3)$
structure manifolds. 
Another useful relation that follows from (\ref{Fierz}) is
\be
\tJ^a_{\;\;b}\tOmega^{bcd}=i\tOmega^{acd}.
\ee

We can connect the various components of the contorsion tensor $\tau_{abc}$ to the
forms that define the $SU(3)$ structure as 
\be \label{tauformulas}
\begin{split}
&\tJ^{ab}\tau_{abc}v^c*1=\half\diff v \wedge v\wedge\tJ\wedge\tJ\\
&\tau_{abc}\tOmega^{abc}*1=-\diff\tJ\wedge\tOmega\wedge v, \\
&v^a(\tau_{abc}-\tau_{bac})\tJ^{bc}*1=\frac{1}{8}(\diff\tOmega\wedge\btOmega
+\diff\btOmega\wedge\tOmega) \, .
\end{split}
\ee
Notice that the $\tau_{abc}(v\wedge \tJ)^{abc}$ term in (\ref{wg}) leads to both
the first and the third term in the expression above. In order to use (\ref{tauformulas}),  
that term is most efficiently decomposed as
\be
\tau_{abc}(\tJ\wedge v)^{abc}
=3\tau_{abc}\tJ^{[ab}v^{c]}
=\tau_{abc}(\tJ^{ab}v^c+\tJ^{bc}v^a+\tJ^{ca}v^b)
=v^a(\tau_{abc}-\tau_{bac})\tJ^{bc}+\tJ^{ab}\tau_{abc}v^c\;.
\ee


\bibliography{FluxRefs}
\bibliographystyle{JHEP}

\end{document}